\documentclass[twocolumn,twoside,slac_two]{revtex4}
\usepackage{graphicx}
\usepackage{fancyhdr}
\RequirePackage{xspace}
\newcommand{\AmS}{{\protect\the\textfont2
  A\kern-.1667em\lower.5ex\hbox{M}\kern-.125emS}}
\usepackage{relsize}
\def\babar{\mbox{\slshape B\kern-0.1em{\smaller A}\kern-0.1em
    B\kern-0.1em{\smaller A\kern-0.2em R}}}
\def\CP                {\ensuremath{C\!P}\xspace}
\def\CPT               {\ensuremath{C\!PT}\xspace} 

\begin{document}

\title{Experimental Prospects for \CP and T Violation Studies in Charm\footnote{University of Cincinnati preprint number UCHEP-07-08}}

%

\author{G. Mancinelli}
\affiliation{University of Cincinnati, Cincinnati, OH 45221, USA
\\
from the \babar\ Collaboration}

\begin{abstract}
We present the current status of experimental results and prospects for
the determination of \CP and T violation in the charm sector. Such
measurements have acquired renewed interest in recent years in view of
theoretical work, which has highlighted the possibility to
probe experimental signatures from New Physics beyond the Standard Model,
since the effect of \CP violation due to Standard Model processes is expected to
be highly 
suppressed in $D$ decays. The current limits of experimental sensitivities
for these studies are reaching the interesting theoretical regimes. 
We include new measurements from the Belle, \babar, and CLEO-c collaborations. 
\end{abstract}

\maketitle

\thispagestyle{fancy}


\section{Introduction}

The amount of \CP violation (CPV) currently discovered in nature is not
sufficient to 
explain the 
universe as we see it. Looking in the charm sector is a  
natural extension of this task. Three are the kinds of CPV we
deal with: CPV in the $D^0-\overline {D^0}$ mixing matrix,  
which is expected to be insignificant in the charm sector, CPV in 
the decay amplitudes, and CPV in the interference between mixing and
decay, which should
be very small as well.  
The second one is also known as direct CPV and will be covered in
this paper. 

The expression for the \CP asymmetry resulting from a process $f$ and its \CP
conjugate $\bar f$ is given by: 

\begin{eqnarray}
A_{CP}&=& \frac{\Gamma(f)-\Gamma(\bar f)}{\Gamma(f)+\Gamma(\bar f)}
 \nonumber \\
&=& \frac{2
  \Im{(A_1A_2^*)}\sin{(\delta_1-\delta_2)}}{|A_1|^2+|A_2|^2+\Re{(A_1A_2^*)}\cos{(\delta_1-\delta_2)}} 
\label{eq1}
\end{eqnarray}

$A_1$ and $A_2$ are two components of the decay amplitude and
$\delta_1-\delta_2$ the corresponding strong phase difference.  
It follows that two amplitudes with different strong as well as weak
phases are needed to have CPV. In the realm 
of the SM, usually this means a tree and a penguin amplitude. 
The kinds of processes described in the following are categorized as
Cabibbo favored (CF, $c\rightarrow s \bar d u$), 
suppressed (CS, $c\rightarrow s \bar s u$, $c\rightarrow d \bar d u$), 
and doubly suppressed (DCS, $c\rightarrow d \bar s u$), according to the
kind of vertices that intervene in the charm quark 
decay.

In contrast to the beauty sector, the Standard Model (SM) charm sector is 
largely \CP conserving, as it
involves 4 quarks and the $2\times2$ Cabibbo mixing matrix is real. 
In singly Cabibbo suppressed decays 
diluted weak phases can produce asymmetries of the order $10^{-3}
-10^{-4}$, while no weak phases, hence no CPV, exist in CF 
and DCS decays, except for some 
minimal asymmetry in the $D^+\rightarrow K_S\pi^+$ mode. 
It is interesting to notice that it is possible in principle to
distinguish direct and indirect CPV, either combining direct \CP 
asymmetries with time-dependent measurements both for 
CP eigenstates, or just using time integrated measurements for CF CP
eigenstate modes (assuming negligible CPV in CF modes) as 
$K_S\pi^0$~\cite{the1}.

New Physics (NP) can contain \CP violating couplings that could show up at
the percent level~\cite{the1}~\cite{the2}~\cite{the3}~\cite{the4}. Several
extensions of the SM predict such asymmetries, 
including models with leptoquarks, a fourth generation of fermions,
right-handed weak currents, or extra Higgs doublets. 
Precision measurements and theory are 
required to detect NP. The charm sector is in a unique position to test
physics beyond the SM.  
In particular it can test models where CPV is generated in
the up-like quark sector. Flavor models where the CKM mixing is 
generated in the up sector generally predict 
large D-mixing and sizable CPV in charm, but smaller effects in
the beauty sector. Furthermore, SCS $D$ decays are now 
more sensitive to gluonic penguin amplitudes 
than are charmless B decays~\cite{the1}. 
In summary, finding CPV in CF and DCS decays or finding CPV
above 0.1\% in SCS decays would indicate NP. 

\section{Current Experimental Results}

\begin{figure}[h]
\centering
\includegraphics[width=80mm]{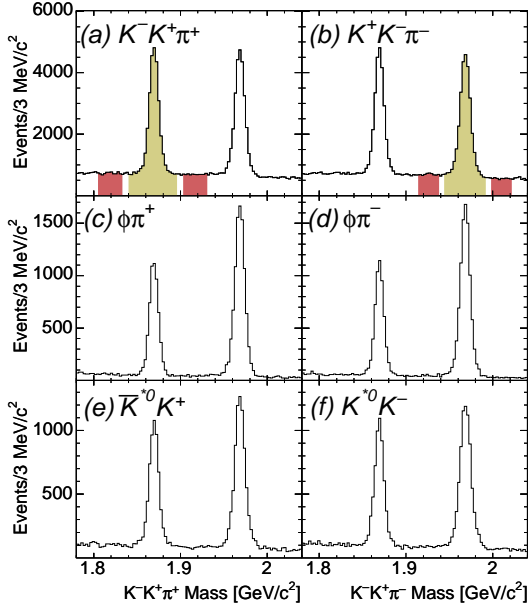}
\caption{\babar's sample of 
$D^+\rightarrow K^-K^+\pi^+$ and $D^+\rightarrow \pi^+\pi^+\pi^-$
candidates (mass distributions) used for CPV measurements.} \label{fig1}
\end{figure}

There are several ways direct \CP and T violation can be looked for: by
measuring asymmetries in time integrated partial widths or in final 
state distributions of Dalitz plots or by measuring 
T violation via T-odd correlations with 4-body $D$ decays.
As an example of charged $D$ decays, Fig.~\ref{fig1} shows the
reconstructed mass distributions in
$D^+\rightarrow K^-K^+\pi^+$ and $D^+\rightarrow \pi^+\pi^+\pi^-$
candidates in the \babar\ detector, with fairly large  
datasets (80 $fb^{-1}$)~\cite{babar1}. 
The \CP asymmetry results of these and many other analyses are listed in
Table~\ref{tab1}.   

As mentioned earlier, in the SM we can expect direct CPV at the $10^{-3}$
level in SCS decays, but no CPV in CF modes. In the $K^+K^-$ and $\pi^+\pi^-$
modes it is puzzling that the ratio  
of their branching ratios is so different from 1 ($\sim$2.8). This
entails the presence of large final state interactions (FSI) and/or large
penguin contributions, which could be fertile  
ground for NP 
to manifest itself. Phenomenological calculations set SM limits at or well
below the $10^{-3}$ level~\cite{bucc}. The CDF collaboration has to date
the best \CP 
measurements for these modes. The asymmetries are
normalized to the CF $K\pi$ mode. Difficulties for this analysis are due
to the track charge asymmetry which is  
calibrated 
with $K_S$ control samples and to the partially reconstructed $D$ background
for the $K^+K^-$ mode. Total systematics are slightly above 0.5$\%$~\cite{cdf}. 

A high precision analysis at \babar\ reports
$A_{CP}=0.00\pm0.34\pm0.13$  for the $K^+K^-$ mode and
$A_{CP}=-0.24\pm0.52\pm0.22$  for the $\pi^+\pi^-$ mode~\cite{babar2}. To
keep the 
systematics so low, the keys are to 
calibrate charge and  
tagging asymmetries using data, namely the CF $K\pi$ mode, and to account
for forward-backward asymmetries due to QED effects, which can produce
detection asymmetries in a
detector as \babar, due to the boost of the center of mass
system with respect to the laboratory.

Charm factories benefit with respect to the beauty ones from a pure
$D\bar D$ final state with low multiplicity, hence high tagging
efficiency. This makes them competitive with the high  
statistics at \babar\ and Belle. Single tag efficiencies range from 25 to
65$\%$, values unimaginable at the B-factories. Most of the new CP
violation results from CLEO-c, with 281 $pb^{-1}$ of data, are for CF  
modes, with the exception of $D^+ \rightarrow K^+K^-\pi^+$ (see
Tables~\ref{tab1} and~\ref{tab2}). 
The uncertainties are of the order of 1$\%$ in most cases, For modes with
charged kaons, the kaon systematics are the largest ones. 
  
In case of indirect CPV and final \CP eigenstates, the time integrated and
time dependent \CP asymmetries are universal and equal to each other. In
contrast, for direct CPV, the time-integrated  
asymmetries in principle are not expected to be universal. Hence parts of
phase-space in a multi-body decay might have different asymmetries (which
may even cancel each other out when 
 integrated over the whole phase-space). In addition, NP might not show up
 in the decay rates asymmetries but instead in the phase difference
 between amplitudes. 3-body decays permit  
the measurement of such phase differences.
The Dalitz plot technique allows increased sensitivity to \CP asymmetry by
probing the decay amplitude rather than the decay rate and access to both
CP eigenstates and non \CP eigenstates with relatively high
statistics. The CLEO-c collaboration has measured the \CP asymmetry in the
$\pi^+\pi^-\pi^0$ 
mode (integrated over the sum of all amplitudes in the Dalitz plot) and
has also  
performed a full fledged Dalitz plot analysis of the $D^0 \rightarrow
K_S\pi^+\pi^-$ decay~\cite{cleoc}. The \babar\ and Belle
collaborations can
exploit their larger datasets for similar measurements. 

T violation measurements can be performed exploiting T-odd correlations
between the momenta of the decay products of 4-body $D$ decays as
$KK\pi\pi$, while assuming \CPT conservation: 

\begin{equation}
C_T=p_{K^+}\cdot(p_{\pi^+}\times p_{\pi^-}).
\label{eq2}
\end{equation}
 
Under time reversal $C_T$ changes sign, but its being different from 0 is not
sufficient to establish T-violation as final state interactions can fake
this asymmetry~\cite{bigi}. To 
overcome this problem the 
analogous quantity from the \CP conjugate decay can be defined as: 
 
\begin{equation}
\overline {C_T}=p_{K^-}\cdot(p_{\pi^-}\times p_{\pi^+}).
\label{eq3}
\end{equation}

Finding $\overline {C_T}\neq -C_T$ establishes T violation. T-odd asymmetries can be built as:
 
\begin{eqnarray}
A_T= & \frac{\Gamma(C_T>0)-\Gamma(C_T<0)}{\Gamma(C_T>0)+\Gamma(C_T<0)} \\ 
\overline {A_T}= & \frac{\Gamma(\overline {C_T}>0)-\Gamma(\overline {C_T}<0)}{\Gamma(\overline
  {C_T}>0)+\Gamma(\overline {C_T}<0)}
\label{eq4}
\end{eqnarray}

and the T violation asymmetry as: 

\begin{equation}
A_{T-viol}=\frac{1}{2} (A_T-\overline {A_T})
\label{eq5}
\end{equation}

and if this is different from 0, T violation is established, even in the
presence of strong phases~\cite{todd}. As for the available measurements the only ones to date are from the FOCUS
collaboration (see Tables~\ref{tab1} and~\ref{tab2}). The CLEO, \babar, and
Belle experiments should better this 
analysis with their larger and 
cleaner data samples. 
   
Most of the measurements of \CP and T violation in neutral $D$ decays to
date are 
shown in Table~\ref{tab1}. No evidence of direct CPV has been found. The best
limits are of the order of one to few  
percent statistical errors with systematics of similar magnitude; few
measurements have errors below the 1$\%$ level. Most of these are old
measurements, except for the new ones  
from the CLEO-c collaboration and the ones ``byproducts'' of the mixing
analyses of the
\babar\ and Belle collaborations.

\begin{table}[h]
\begin{center}
\caption{$A_{CP}$ measurements to date using neutral $D$ decays. The last
  row reports a measurement of $A_T$ by the FOCUS collaboration. }
\begin{tabular}{|l|c|c|}
\hline \textbf{Experiment(year)} & \textbf{Decay mode} & \textbf{$A_{CP}$\%}
\\
\hline
CDF(2005)         & $D^0\rightarrow K^+K^-$                  & $2.0  \pm  1.2  \pm  0.6$  \\
\hline
CLEO(2002)        & $D^0\rightarrow K^+K^-$                  & $0.0  \pm  2.2  \pm  0.8$  \\
\hline
FOCUS(2000)       & $D^0\rightarrow K^+K^-$                  & $-0.1  \pm  2.2  \pm  1.5$  \\
\hline
CDF(2005)         & $D^0\rightarrow \pi^+\pi^-$              & $1.0  \pm  1.3  \pm  0.6$  \\
\hline
CLEO(2002)        & $D^0\rightarrow \pi^+\pi^-$              & $1.9  \pm  3.2  \pm  0.8$  \\
\hline
FOCUS(2000)       & $D^0\rightarrow \pi^+\pi^-$              & $4.8  \pm  3.9  \pm  2.5$  \\
\hline
CLEO(2001)        & $D^0\rightarrow K^0_SK^0_S$              & $-23  \pm  19 $  \\
\hline
CLEO(2001)        & $D^0\rightarrow \pi^0\pi^0$              & $0.1  \pm  4.8$  \\
\hline
CLEO(2001)        & $D^0\rightarrow K^0_S\pi^0$              & $0.1  \pm  1.3$  \\
\hline
CLEO(1995)        & $D^0\rightarrow K^0_S\phi$               & $2.8  \pm  9.4$  \\
\hline
CLEO(2005)        & $D^0\rightarrow \pi^+\pi^-\pi^0$         & $1  ^{+9}_{-7}  \pm  5$  \\
\hline
CLEO(2004)        & $D^0\rightarrow K^0_S\pi^+\pi^-$         & $-0.9  \pm  2.1  ^{+1.6}_{-5.7}$  \\
\hline
Belle(2005)       & $D^0\rightarrow K^+\pi^+\pi^-\pi^-$      & $-1.8  \pm  4.4$ \\
\hline
FOCUS(2005)       & $D^0\rightarrow K^+K^-\pi^+\pi^-$        & $-8.2  \pm  5.6  \pm  4.7$  \\
\hline
CLEO(2007)        & $D^0\rightarrow K^-_S\pi^+$              & $-0.4  \pm  0.5  \pm  0.9$  \\
\hline
CLEO(2007)        & $D^0\rightarrow K^-_S\pi^+\pi^0$         & $0.2  \pm  0.4  \pm  0.8$  \\
\hline
CLEO(2007)        & $D^0\rightarrow K^-\pi^+\pi^+\pi^-$      & $0.7  \pm  0.5  \pm  0.9$  \\
\hline
Belle(2005)       & $D^0\rightarrow K^+\pi^-\pi^0$           & $-0.6  \pm  5.3 $  \\
\hline
\babar(2007)       & $D^0\rightarrow K^+\pi^-$                & $-2.1  \pm  5.2  \pm  1.5$  \\
\hline
Belle(2007)       & $D^0\rightarrow K^+\pi^-$                & $2.3  \pm  4.7  $  \\
\hline
FOCUS(2005) $A_T$       & $D^0\rightarrow K^+K^-\pi^-\pi^+$        & $1.0  \pm  5.7  \pm  3.7$  \\
\hline
\end{tabular}
\label{tab1}
\end{center}
\end{table}

Table~\ref{tab2} reports the results for charged $D$ decays, where many
new results 
from CLEO-c in SCS $D_s$ decays and in several CF decays are listed.  
With the CLEO-c result, together with the ones from \babar\ and FOCUS, the 
$KK\pi$ mode is becoming one of the most interesting precision-wise. 
Table~\ref{tab3} lists the average \CP asymmetry measurements by mode. Some
of the 
averages are from HFAG group~\cite{hfag}, some are my own, hence probably
not as 
correct, as not all correlations are taken into account. Clearly  
there is still work to do.

\begin{table}[h]
\begin{center}
\caption{$A_{CP}$ measurements to date using charged $D$ decays. The last
  two rows report measurements of $A_T$ by the FOCUS collaboration.}
\begin{tabular}{|l|c|c|}
\hline \textbf{Experiment(year)} & \textbf{Decay mode} & \textbf{$A_{CP}$\%}
\\
\hline
\babar(2005)         & $D^+\rightarrow K^+K^-\pi^+$    & $ 1.4 \pm 1.0  \pm 0.8$  \\
\hline 
\babar(2005)         & $D^+\rightarrow \phi\pi^+$      & $ 0.2 \pm 1.5  \pm 0.6$  \\
\hline 
\babar(2005)         & $D^+\rightarrow K_S^0K^+$       & $ 0.9 \pm 1.7  \pm 0.7$  \\
\hline 
CLEO(2007)         & $D^+\rightarrow K^+K^-\pi^+$     & $ -0.1 \pm 1.5  \pm 0.8$  \\
\hline 
FOCUS(2000)         & $D^+\rightarrow K^+K^-\pi^+$    & $ 0.6 \pm 1.1  \pm 0.5$  \\
\hline 
E791(1997)         & $D^+\rightarrow K^+K^-\pi^+$     & $ -1.4 \pm 2.9 $  \\
\hline 
E791(1997)         & $D^+\rightarrow \phi\pi^+$       & $ -2.8 \pm 3.6 $  \\
\hline 
E791(1997)         & $D^+\rightarrow K_S^0K^+$        & $ -1.0 \pm 5.0 $  \\
\hline 
FOCUS(2002)         & $D^+\rightarrow K_S^0\pi^+$     & $ -1.6 \pm 1.5  \pm 0.9$  \\
\hline 
CLEO(2007)         & $D^+\rightarrow K_S^0\pi^+$      & $ -0.6 \pm 1.0  \pm 0.3$  \\
\hline 
CLEO(2007)         & $D^+\rightarrow K_S^0\pi^+\pi^0$       & $ 0.3 \pm 0.9  \pm 0.3$  \\
\hline 
CLEO(2007)         & $D^+\rightarrow K_S^0\pi^+\pi^+\pi^-$  & $ 0.1 \pm 1.1  \pm 0.6$  \\
\hline 
CLEO(2007)         & $D^+\rightarrow K^-\pi^+\pi^+$         & $ -0.5 \pm 0.4  \pm 0.9$  \\
\hline 
CLEO(2007)         & $D^+\rightarrow K^-\pi^+\pi^+\pi^0$    & $ 1.0 \pm 0.9  \pm 0.9$  \\
\hline 
CLEO(2007)         & $D_S^+\rightarrow K^+\eta$             & $ -20 \pm 18$  \\
\hline 
CLEO(2007)         & $D_S^+\rightarrow K^+\eta'$            & $ -17 \pm 37$  \\
\hline 
CLEO(2007)         & $D_S^+\rightarrow K_S^0\pi^+$          & $ 27 \pm 11 $  \\
\hline 
CLEO(2007)         & $D_S^+\rightarrow K^+\pi^0$            & $ 2 \pm 29 $  \\
\hline 
E791(1997)         & $D^+\rightarrow \pi^+\pi^+\pi^-$       & $ -1.7 \pm 4.2  $  \\
\hline 
FOCUS(2005) $A_T$        & $D^+\rightarrow K_S^0K^+\pi^+\pi^-$   & $ 2.3 \pm 6,2  \pm 2.2$  \\
\hline
FOCUS(2005) $A_T$        & $D_S^+\rightarrow K_S^+K^+\pi^+\pi^-$ & $ -3.6 \pm 6.7  \pm 2.3$  \\
\hline
\end{tabular}
\label{tab2}
\end{center}
\end{table}

\begin{table}[h]
\begin{center}
\caption{Average \CP asymmetry measurements by mode.}
\begin{tabular}{|l|c|}
\hline \textbf{Decay mode} & \textbf{$A_{CP}$\%}
\\
\hline
$D^0\rightarrow K^+K^-$            & $ +1.4 \pm 1.2 $  \\
\hline 
$D^0\rightarrow K_S^0K_S^0$        & $ -2.3 \pm 1.9 $  \\
\hline 
$D^0\rightarrow \pi^+\pi^-$        & $ +1.3 \pm 1.3 $  \\
\hline 
$D^0\rightarrow \pi^0\pi^0$        & $ +0.1 \pm 4.8 $  \\
\hline 
$D^0\rightarrow \pi^+\pi^-\pi^0$   & $ +1   \pm 9   $  \\
\hline 
$D^0\rightarrow K_S^0\pi^0$        & $ +0.1 \pm 1.3 $  \\
\hline 
$D^0\rightarrow K^-\pi^+$          & $ -0.4 \pm 1.0 $  \\
\hline 
$D^0\rightarrow K^-\pi^+\pi^0$     & $ +0.2 \pm 0.9 $  \\
\hline 
$D^0\rightarrow K^-\pi^+\pi^+\pi^-$    & $ +0.7 \pm 1.0 $  \\
\hline 
$D^0\rightarrow K^+\pi^-$              & $ -0.8 \pm 3.1 $  \\
\hline 
$D^0\rightarrow K^+\pi^-\pi^0$         & $ -0.1 \pm 5.2 $  \\
\hline 
$D^0\rightarrow K+S^0\pi^+\pi^0$       & $ -0.9 \pm 4.2 $  \\
\hline 
$D^0\rightarrow K^+\pi^-\pi^+\pi^-$    & $ -1.8 \pm 4.4 $  \\
\hline 
$D^0\rightarrow K^+K^-\pi^-\pi^+$      & $ -8.2 \pm 7.3 $  \\
\hline 
$D^+\rightarrow K_S^0\pi^+$            & $ -0.9 \pm 0.9 $  \\
\hline
$D^+\rightarrow K_S^0\pi^+\pi^0$       & $ +0.3 \pm 0.9 $  \\
\hline
$D^+\rightarrow K_S^0\pi^+\pi^+\pi^-$  & $ +0.1 \pm 1.3 $  \\
\hline
$D^+\rightarrow K^-\pi^+\pi^+$         & $ -0.5 \pm 1.0 $  \\
\hline
$D^+\rightarrow K^-\pi^+\pi^+\pi^0$    & $ +1.0 \pm 1.3 $  \\
\hline
$D^+\rightarrow K_S^0K^+$               & $ +7.1 \pm 6.2 $  \\
\hline
$D^+\rightarrow K^+K^-\pi^+$           & $ +0.6 \pm 0.8 $  \\
\hline
$D^+\rightarrow \pi^+\pi^-\pi^+$       & $ -1.7 \pm 4.2 $  \\
\hline
$D^+\rightarrow K_S^0K^+\pi^+\pi^-$    & $ -4.2 \pm 6.8 $  \\
\hline
\end{tabular}
\label{tab3}
\end{center}
\end{table}

\section{Future Prospects}

The future prospects for these measurements are very promising. For the
$KK$ and $\pi\pi$ modes both B-factories and CDF are expected to reach
very interesting sensitivities of the order of few per thousand. The issue at
the Tevatron will be whether the trigger can cope with the increase of
luminosity. The $D^+ \rightarrow KK\pi$ mode should hit interesting limits
as well, if systematics can be hold under control. Very promising are also
measurements from  Dalitz plot analyses using SCS modes, where we have the
added puzzle that it is not known where (if anywhere) CPV can show up in
the Dalitz plane. Furthermore, the asymmetry could be large, but
confined  to only a part of the phase-space.  

For the T-correlation analyses, as aforementioned there are large
datasets of 4-body $D$ decays available. The \babar\ and Belle
collaborations could achieve
statistical uncertainties at or below 0.5$\%$ if systematics can
be kept as low. 

If the present machines cannot fully probe the extent of the CP
asymmetries allowed by NP or given to us by nature, new
experiments at BESIII and at the super B-factories at KEK and/or Frascati,
or LHC-b at the LHC definitely will. Data taking at BESIII is expected
to start in 
2008. With 3 years of running it will get up to 20 times the CLEO-c
dataset. The super B-factories should record $\sim$10 $ab^{-1}$ of data per
year. At least the super B-factory at Frascati is designed to run at the
psi(3770) as well with an estimated 1 $ab^{-1}$ of data per year. LHC-b
will implement a dedicated $D^*$ trigger as well, selecting huge and
clean samples of hadronic $D$ decays. This should assure a $D^*$ dataset of
the order of 
100 times the one at CDF in the first year of nominal luminosity
running. With all this data we will reach a phase of high precision CPV
measurements, with uncertainties of the order of $10^{-4}$ ($6\times
10^{-5}$) with one year of nominal running at LHC-b (super-B factories). 

\section{Conclusions}

Charm physics provides unique opportunities for indirect
searches for new physics. The theoretical calculations of the $D^0$ mixing
parameters have
large uncertainties, hence physics beyond  
the standard model will be hard to rule out from $D^0$ mixing measurements
alone. The observation of large CPV would instead be a clear and robust
signal of new physics. There have been some exciting new results this
year from the CLEO-c, 
Belle, and \babar\ collaborations. The total  
uncertainties are at the 1$\%$ level in several modes, but still far from
observation. Experiments are just now entering the interesting
domain. The future ahead is very promising with good  
sensitivities achievable by current experiments and high precision
measurements expected with future and planned efforts.  

\begin{acknowledgments}

I would like to thank the conference organizers  for their
help and gracious hospitality in Ithaca, my charming mentors, Brian
Meadows and Mike Sokoloff, and students, Kalanand Mishra and Rolf
Andreassen, from the University of Cincinnati, as well as all my other
colleagues who have been, are, and 
will be working on this exciting subject. This work was supported in part
by the National Science 
Foundation Grant Number PHY-0457336 and by the Department of Energy Contract
Number DE-AC03-76SF00515. 

\end{acknowledgments}

\bigskip 


\end{document}